\documentclass[twocolumn,prb,aps,floatfix]{revtex4-1}
\usepackage{bm,color,amsmath,amssymb,mathrsfs,dsfont,latexsym,graphicx,psfrag,float}
\usepackage[caption=false]{subfig}
\usepackage{times}

\newcommand{\be}{\begin{equation}}
\newcommand{\ee}{\end{equation}}
\newcommand{\bea}{\begin{eqnarray}}
\newcommand{\eea}{\end{eqnarray}}

\renewcommand{\phi}{\varphi}
\renewcommand{\epsilon}{\varepsilon}
\renewcommand{\vec}[1]{{\bf #1}}

\begin{document}
\fontsize{.32cm}{.34cm}\selectfont

\title{Topolectrical circuits}
\author{Ching Hua Lee} 
\affiliation{Institute of High Performance Computing, A*STAR,
  Singapore, 138632}
	\affiliation{Department of Physics, National University of Singapore, Singapore, 117542}

\author{Stefan Imhof} 
\affiliation{Experimentelle Physik 3, Physikalisches Institut,
  University of W\"urzburg, Am Hubland, D-97074 W\"urzburg, Germany}

\author{Christian Berger} 
\affiliation{Experimentelle Physik 3, Physikalisches Institut,
  University of W\"urzburg, Am Hubland, D-97074 W\"urzburg, Germany}

\author{Florian Bayer} 
\affiliation{Experimentelle Physik 3, Physikalisches Institut,
  University of W\"urzburg, Am Hubland, D-97074 W\"urzburg, Germany}

\author{Johannes Brehm} 
\affiliation{Experimentelle Physik 3, Physikalisches Institut,
  University of W\"urzburg, Am Hubland, D-97074 W\"urzburg, Germany}

\author{Laurens W. Molenkamp} 
\affiliation{Experimentelle Physik 3, Physikalisches Institut,
  University of W\"urzburg, Am Hubland, D-97074 W\"urzburg, Germany}

\author{Tobias Kiessling} 
\affiliation{Experimentelle Physik 3, Physikalisches Institut,
  University of W\"urzburg, Am Hubland, D-97074 W\"urzburg, Germany}

\author{Ronny Thomale} 
\email{Corresponding author: rthomale@physik.uni-wuerzburg.de}
\affiliation{Institute for Theoretical Physics and Astrophysics,
  University of W\"urzburg, Am Hubland, D-97074 W\"urzburg, Germany}

\date{\today}

\begin{abstract}
\end{abstract}


\maketitle

\textbf{Invented by Alessandro Volta and F\'elix Savary in the early 19th century, circuits consisting of resistor, inductor and capacitor (RLC) components are  omnipresent in modern technology. The behavior of an RLC circuit is governed by its circuit Laplacian, which is analogous to the Hamiltonian describing the energetics of a physical system. We show that topological semimetal band structures can be realized as admittance bands in a periodic RLC circuit, where we employ the grounding to adjust the spectral position of the bands similar to the chemical potential in a material.
Topological boundary resonances (TBRs) appear in the impedance read-out of a topolectrical circuit, providing a robust signal for the presence of topological admittance bands. For experimental illustration, we build the Su-Schrieffer-Heeger circuit, where our impedance measurement detects a TBR related to the midgap state.
Due to the versatility of electronic circuits, our topological semimetal construction can be generalized to band structures with arbitrary lattice symmetry. Topolectrical circuits establish a bridge between electrical engineering and topological states of matter, where the accessibility, scalability, and operability of electronics synergizes with the intricate boundary properties of topological phases.}

Topological semimetals~\cite{anton} constitute the latest development of an evolution dating back more than thirty years, when topological phases began to cast their shadows before as midgap states in polyacetylene~\cite{su-79prl1698} and the quantized edge modes of integer quantum Hall systems~\cite{klitzing-80prl494} were discovered. 
Driven by the flourishing field of topological insulators~\cite{RevModPhys.82.3045,RevModPhys.83.1057}, the viewpoint of topology has recently branched out to various classes of physical systems, ranging from electrons in solids to photonic networks in metamaterials, ultra-cold atoms in optical lattices, microwave resonators, electrical circuits, and phonons in mechanical setups 
(see e.g. Refs.~\onlinecite{marin,lubensky,ningyuan2015time,susstrunk2015observation,yang2015topological,budich}). Note that such topological states of matter do not necessarily rely on any quantum mechanical framework. In mathematical terms, it is not the quantum, i.e. non-commutative, nature of the Hilbert space, but rather the non-trivial connectivity of phase space under cyclic evolution of parameters~\cite{Berry45} that indicates a topological phase. 

The fingerprint of a topological insulator motif, independent of the physical setting in which it is realized, is given by a single edge mode response protected by topology, along with an unresponsive bulk. While there are various promising approaches to realize them within classical arrays, 
topological device design is often limited due to insufficient edge mode density. As opposed to fermionic systems where the chemical potential is a useful parameter to access any particular range of the band structure at low energies, bosonic or classical degrees of freedom for a topological band structure also pose the problem how to systematically address the spectral regime of interest, such as the band gap domain of a topological insulator.   
Furthermore, in an era where classical experimental setups for topological phases still need to improve in terms of uniformity of array elements, it is often challenging to resolve single edge mode responses to identify the onset of a topological insulator phase. 


We propose the topological semimetal paradigm in classical RLC circuits, which predicts highly pronounced resonances in a generic impedance read-out whenever there are topological boundary modes that scale extensively, such as the Fermi arcs of topological semimetals~\cite{wan2011topological}. Due to their extensive degeneracy, such topological boundary resonances (TBRs) remain robust even in the face of significant nonuniformity of circuit elements, promising high-precision identification in a realistic measurement. We outline a detailed design of such topolectrical circuits, including a Weyl circuit network exhibiting TBRs of Fermi arc type, where the AC driving frequency combined with the grounding design takes over the role of the chemical potential in a fermionic system. As an initial proof-of-principle experimental study, we report impedance and voltage profile measurements of the Su-Schrieffer-Heeger circuit chain. As a theoretical byproduct in this work, we further introduce the mathematical framework for characterizing topological properties of electrical circuit graphs in general. While our semimetal paradigm can be applied to any classical array setup such as mechanical systems or optical cavities, the topolectrical circuits we introduce combine all desired conceptual and experimental preferences to realize topological semimetal analogs in a classical model, without demanding specialized equipment.

\begin{figure}
\begin{minipage}{\linewidth}
\includegraphics[width=0.99\linewidth]{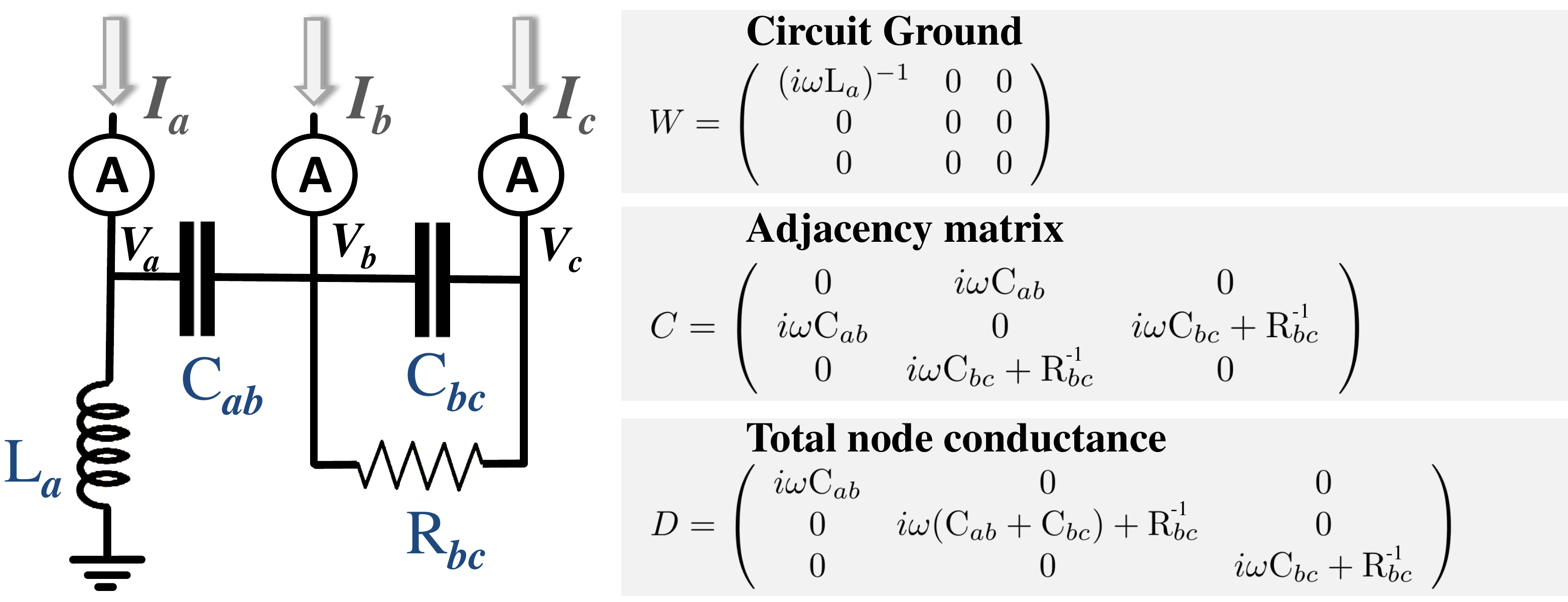}
\end{minipage}
\caption{Definition of the principal building blocks for the grounded circuit Laplacian $J=D-C+W$ (Eq. 2) of an illustrative RLC circuit with nodes $\{a,b,c\}$. $W$ and $D$ are diagonal matrices containing the total conductances from each node towards the ground and towards the rest of the circuit, respectively. $C$ is the adjacency matrix of the circuit graph, with edges weighted by their conductances. 
}
\label{topocircuit}
\end{figure}

Any electrical circuit network can be represented by a graph whose nodes and edges correspond to the circuit junctions and connecting wires/elements. The circuit behavior is fundamentally described by Kirchhoff's law 
\begin{equation}
I_a=\sum_{i}\text{C}_{ai}(V_a-V_i)+w_aV_a,
\label{LV0}
\end{equation}
where $I_a$ and $V_a$ are the input current and electrical potential at each node $a$. 
By current conservation, $I_a$ equals the total current flowing out of node $a$ towards all other nodes $i$ linked by nonzero conductance $\text{C}_{ai}$, plus the current flowing into the ground through a route with impedance $w_a^{-1}$. 
The impedance and conductances are real for resistive circuit elements, but will be complex when capacitors or inductors are present (Fig. 1).
As an initial step towards identifying circuits with tight-binding lattice models, we  rewrite Eq.~\ref{LV0} in compact matrix form 
\begin{equation}
\vec I=(L+W)\vec V := J\vec V ,
\label{LV}
\end{equation}
with vectors $\vec V$ and $\vec I$ formed by the components $V_a$ and $I_a$. The grounded Laplacian $J$ consists of $L$, the circuit Laplacian which depends on the conductance network structure, and $W=\text{diag}(w_1,w_2,\dots)$, which depends on how the circuit is grounded. 
The Laplacian is defined in terms of the conductances by $L=D-C$, 
where $C$ is the (adjacency) matrix of conductances and $D=\text{diag}(\sum_{i}\text{C}_{1i}, \sum_{i}\text{C}_{2i},\dots)$ lists the total conductances out of each node (Fig. 1). To understand the relation of $L$ with the continuum Laplacian, one writes the spreading of current from a node as a divergence of current density $I=\nabla\cdot \vec j$, and invokes the definition of conductivity $\vec j=\sigma\vec E=\sigma\nabla V$. Hence $\vec I=\nabla\cdot (\sigma\nabla)\vec V=L\vec V$. This establishes $L$ as the continuum Laplacian restricted to a circuit. 

\begin{widetext}
\begin{figure}
\begin{minipage}{.9\textwidth}
\includegraphics[width=\linewidth]{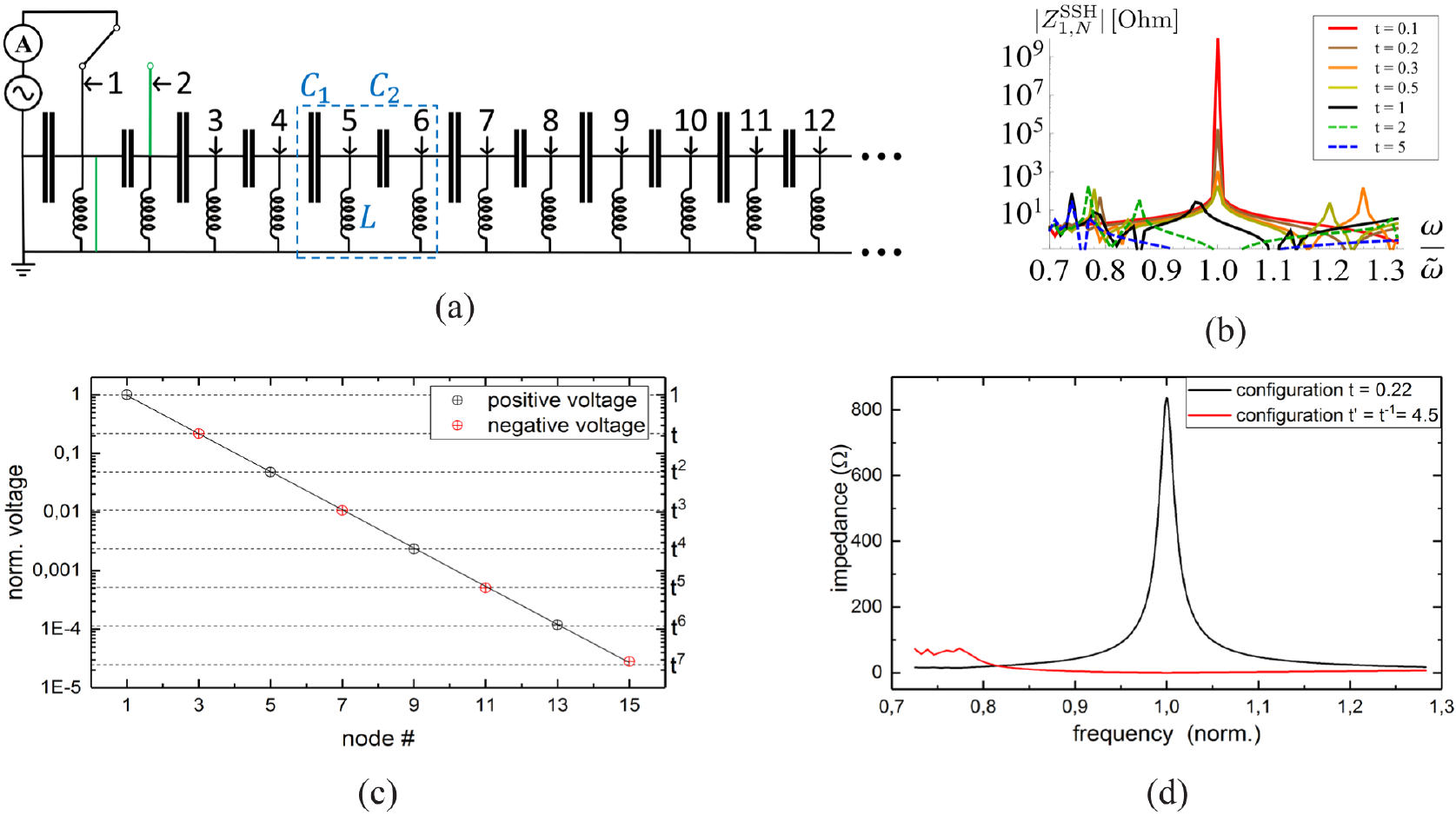}
\caption{(a) Circuit diagram of the SSH topolectrical circuit. Each unit cell consists of a pair of capacitors $C_1$ and $C_2$, with identical inductors $L$ between every two capacitors. An AC source 
provides a driving voltage with amplitude $V_0$. For $t=C_1/C_2<1$, an SSH midgap mode is found. In the experimental implementation we set $C_1 = 0.1 \mu$F, $C_2=0.22 \mu$F and L=$10 \mu$H. Green lines indicate wiring for measuring the $t^{-1}$-configuration on the same circuit. 
(b) Ideal impedance magnitude across the nodal ends $a=1$ and $b=N$ of an $N=10$ SSH topolectrical circuit as a function of AC frequency $\omega$ for various values of $t$. The dashed curves highlight topologically trivial cases for $t>1$, showing that the impedance increases enormously only for $t<1$, the topologically nontrivial regime. The TBR at $\omega=\tilde \omega$ is most pronounced for the smallest $t$, and decreases exponentially as $t$ is increased. Secondary resonances are observed at larger deviations from $\tilde \omega$, and are associated with other eigenvalues of the grounded circuit Laplacian $J$. (c) Measurement of $\psi_0(n)$, which accurately fits the shape predicted by theory, i.e., $\psi_0(n)=((-t)^{n}V_0,0)$ for the $n$th two-site unit cell from the left, see also (a) for node numbering.
(d) Impedance measurement of the $t=0.22$ and $t^{-1}=4.5$ configuration.  Despite non-negligible serial resistances and element non-uniformities, the SSH midgap peak is clearly observed in the impedance measurement and absent for the $t^{-1}=4.5$ configuration.}
\label{circuit1D}
\end{minipage}
\end{figure}
\end{widetext}

A circuit is most commonly studied through an impedance measurement, which involves running a current through it and measuring the voltage response. As capacitive and inductive resistances explicitly depend on it, the driving voltage frequency $\omega$ is a central tuning parameter of topolectrical circuits. The simplest measurement is the two-point impedance $Z_{ab}=(V_a-V_b)/I$ between nodes $a$ and $b$, where $V_a-V_b$ is their potential difference and $I$ is the magnitude of the current $I_{a,b}=\pm I$
that enters at $a$ and leaves at $b$. To determine $Z_{ab}$, the potentials have to be expressed in terms of the input current by inverting Eq. \ref{LV}. For this purpose, we employ the regularized inverse of $J$ known as the
circuit Green's function $G=\sum_{j_n\neq 0} \frac1{j_n}\psi^{\phantom{\dagger}}_n \psi_n^\dagger$, 
where $j_n$ and $\psi_n$ denote the admittance eigenvalues and the $N$-dimensional eigenmode vectors of $J$, respectively.
(Regularization in this context means that $j_n=0$ modes are omitted when the circuit is not grounded ($W=0$) and hence defined up to an overall potential offset. If $J$ is not Hermitian, $\psi_n^\dagger$ and $\psi_n$ are replaced by the left and right eigenvectors.) The eigenmodes are potential distributions proportional to the input current distribution. 
Note that $G$ is always symmetric when the circuit elements are reciprocal (see also Ref.~\onlinecite{vcervnanova2014non}). 
The two-point impedance reads~\cite{cserti2011uniform}
\begin{eqnarray}
Z_{ab}&=&\sum_{i=a,b}\frac{G_{ai}I_i-G_{bi}I_i}{I}=G_{aa}+G_{bb}-G_{ab}-G_{ba} \notag\\
&=&\sum_{j_n\neq 0}\frac{|\psi_{n,a}-\psi_{n,b}|^2}{j_n},
\label{Z2}
\end{eqnarray}
where  $\psi_{n,a}-\psi_{n,b}$ is the difference between the amplitudes of the $n$th admittance eigenmode. 
As such, the impedance for each mode $n$ depends on the squared magnitude of its potential difference between $a$ and $b$, weighted by its eigen-impedance $j_n^{-1}$. 


To make contact with topological bandstructures, we consider circuits made up of periodic sublattices.
A node $a=(\vec x,s)$ can be indexed by its unit cell position $\vec x$ and sublattice label $s$. Due to translation symmetry, Bloch's theorem allows us to index the eigenmodes by momentum $\vec k$ and band index $m$, i.e. $\psi_{(\vec k,m)}(\vec x,s)=\phi_m(\vec k,s) e^{i\vec k\cdot \vec x}$.
 Henceforth, we shall call the set of eigenvalues $j_{\vec k,m}$ the bandstructure of the circuit, and also refer to the nodes as sites. The impedance between two sites $(\vec 0,s)$ and $(\vec x,s')$ takes the form
\begin{equation}
Z^{ss'}_{\vec x}=\sum_{\vec k,m}\frac{|\phi_{m}(\vec k,s)-\phi_{m}(\vec k,s')e^{i\vec k\cdot \vec x}|^2}{j_{\vec k,m}},
\label{Z3}
\end{equation}
which reduces to $\sum_{\vec k,m}4j_{\vec k,m}^{-1}|\phi_{m}(\vec k,s)|^2\sin^2\left(\frac{\vec k\cdot \vec x}{2}\right)$ for nodes on the same sublattice $s=s'$. 
The impedance between two nodes becomes large if there exists a finite density of nontrivial eigenmodes with small $j_{\vec k,m}$. 
Such divergences correspond to resonances in RLC circuits, and will be even more pronounced if the relevant eigenmodes are localized at one region, e.g. a boundary of the circuit or a domain wall trajectory. This is the case for TBRs in topolectrical circuits, where there exists a large density of protected boundary modes with $j_{\vec k,m}\approx 0$. A central result of our work will be the construction of such topolectrical circuits with ``grounded'' RLC networks, with the ground controlling the pinning of the TBR to $j_{\vec k,m}\approx 0$. 



The most elementary 2-band topolectrical circuit can be built from a line of capacitors with alternating capacitances of $C_1$ and $C_2$ (Fig. \ref{circuit1D}), which is characterized by $t:=C_1/C_2$. Note that as will be relevant in the following, changing the initial capacitor to the left from $C_1 \rightarrow C_2$ implies $t \rightarrow 1/t$. Identical inductors $L$ connect the junctions between each capacitor to a common isolated grounding plate. For $t<1$, a topological boundary mode exists and leads to a drastic increase in circuit impedance, i.e., a TBR.
Consider one setup of Fig.~\ref{circuit1D}, with the leftmost grounded capacitor of capacitance $C_1<C_2$, and another setup with $C_{1,2}$ interchanged. 
To see that the former arrangement supports a localized "midgap'' eigenmode (configuration of potentials) that decays exponentially to the right, while the latter does not, notice that a fixed amount of charge $Q$ between any pair of $C_1,C_2$ capacitors leads to potential differences $V_1,V_2$ related by $Q=C_1V_1=C_2V_2$ between their respective plates. For $t<1$, there will be a larger potential difference between the plates of $C_1$ than that of $C_2$. Indeed, when driven by an AC supply, $V_1$ and $V_2$ oscillate in anti-phase with relative amplitude $V_2/V_1=t$, corresponding to the potential configuration $\psi_{0}(n)\propto (1,0,-t,0,t^2,0,-t^3,0,...,(-t)^{2n+1})$, where the index $n$ runs through all nodes. $\psi_0(n)$ is exponentially localized at the left end, with a decay length of $\xi=(\log\frac{C_2}{C_1})^{-1}=-\log t$. Since $\vec V$ and the source/sink current $\vec I$ vanish on the even nodes and are proportional to $(-t)^{2n+1}$ on the odd nodes, it follows that $\psi_0 \equiv \vec V \propto \vec I$, i.e., $\psi_0$ is an eigenmode of $J$. 

In the language of the grounded circuit Laplacian, the 
system with periodic boundary conditions is described by
\begin{eqnarray}
J_{\text{SSH}}(k_x)&=&i\omega\left(C_1+C_2-\frac1{\omega^2L}\right)\mathbb{I}\notag\\
&& -i\omega\left[(C_1+C_2\cos k_x)\sigma_x+C_2\sin k_x\sigma_y\right]
\label{J1}
\end{eqnarray}
which, up to prefactors, is equivalent to the enigmatic Su-Schriffer-Heeger (SSH) model developed for midgap states in polyacetylene~\cite{su-79prl1698}. 
Here, $\sigma_x$ and $\sigma_y$ are the Pauli matrices defined in the basis consisting of a $C_1$ capacitor and an adjacent $C_2$ capacitor on its right. The boundary mode $\psi_{0}(x)$, where the notation $x$ is now highlighting the site instead of the node interpretation  $n$, is the circuit analog of the SSH zero mode consisting of ``dimerized'' pairs of capacitors with large amounts of charge oscillating between them. It is topologically protected by a 1D winding number (cf. appendices). 
For $t<1$, one finds a nonzero topological winding
which cannot be deformed into a trivial winding unless the gap, i.e., the spectral gap of the circuit Laplacian, closes. 

Since the left end of the circuit by itself always marks the transition to a trivial regime, for $t<1$ we expect a boundary mode with vanishing spectral value $j_0$ in the semi-infinite limit. Indeed, as shown in the appendix, $j_0\sim(-t)^{N}$, where $N$ denotes the total number of capacitors. This vanishing eigenvalue marks the TBR, which for open boundary conditions and a hypothetically ideal circuit without serial resistance is characterized by a divergent impedance of $Z_{ab}^{\text{SSH}}\sim \frac{(t^{d_a}-t^{d_b})^2}{i\tilde{\omega} C_2(-t)^{N}}$ (Fig.~\ref{circuit1D}b), where $\tilde{\omega}$ denotes the resonant frequency $\tilde{\omega}=1/\sqrt{L(C_1+C_2)}$, and $d_a,d_b$ are the unit cell distances of nodes $a$ and $b$ from the leftmost capacitor. 

The theoretical prediction described above is rather precisely what we find experimentally. In the setup depicted in Fig.~\ref{circuit1D}a, we can switch between a capacitor ratio of $t$ and $1/t$ depending on how fix the switch to node 1 or 2, which affects the boundary condition where the external voltage source is applied. No midgap mode at the external voltage frequency $\tilde{\omega}$ is observed for $t>1$, but for $t<1$. For experimental convenience, we have constrained ourselves to measuring the impedance at the pair of nodes at the boundary, and further map out the midgap voltage profile eigenstate $\psi_{0}(n)$ of the circuit by measuring the voltage difference between neighboring nodes (Fig.~\ref{circuit1D}c,d). $\psi_0 (n)$ displays the predicted behaviour within negligible error bars. We find the latter to be a robust measurement, along with the predicted impedance profile if we allow for non-uniformity of circuit elements and consider serial circuit resistance in our calculations (cf. appedices). 





\begin{figure}
\begin{minipage}{\linewidth}
\subfloat[]{\includegraphics[width=0.9\linewidth]{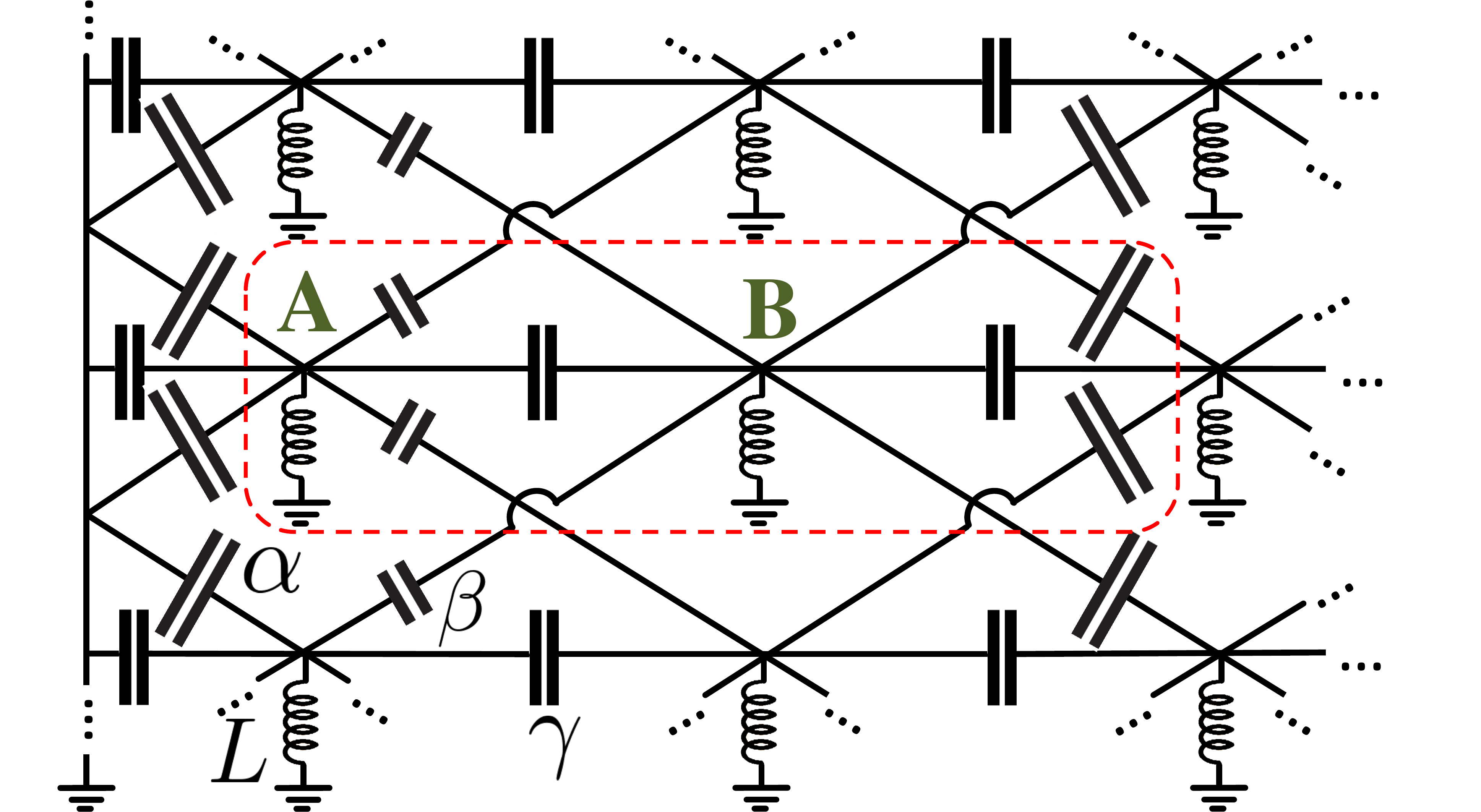}}\\
\subfloat[]{\includegraphics[width=0.44\linewidth]{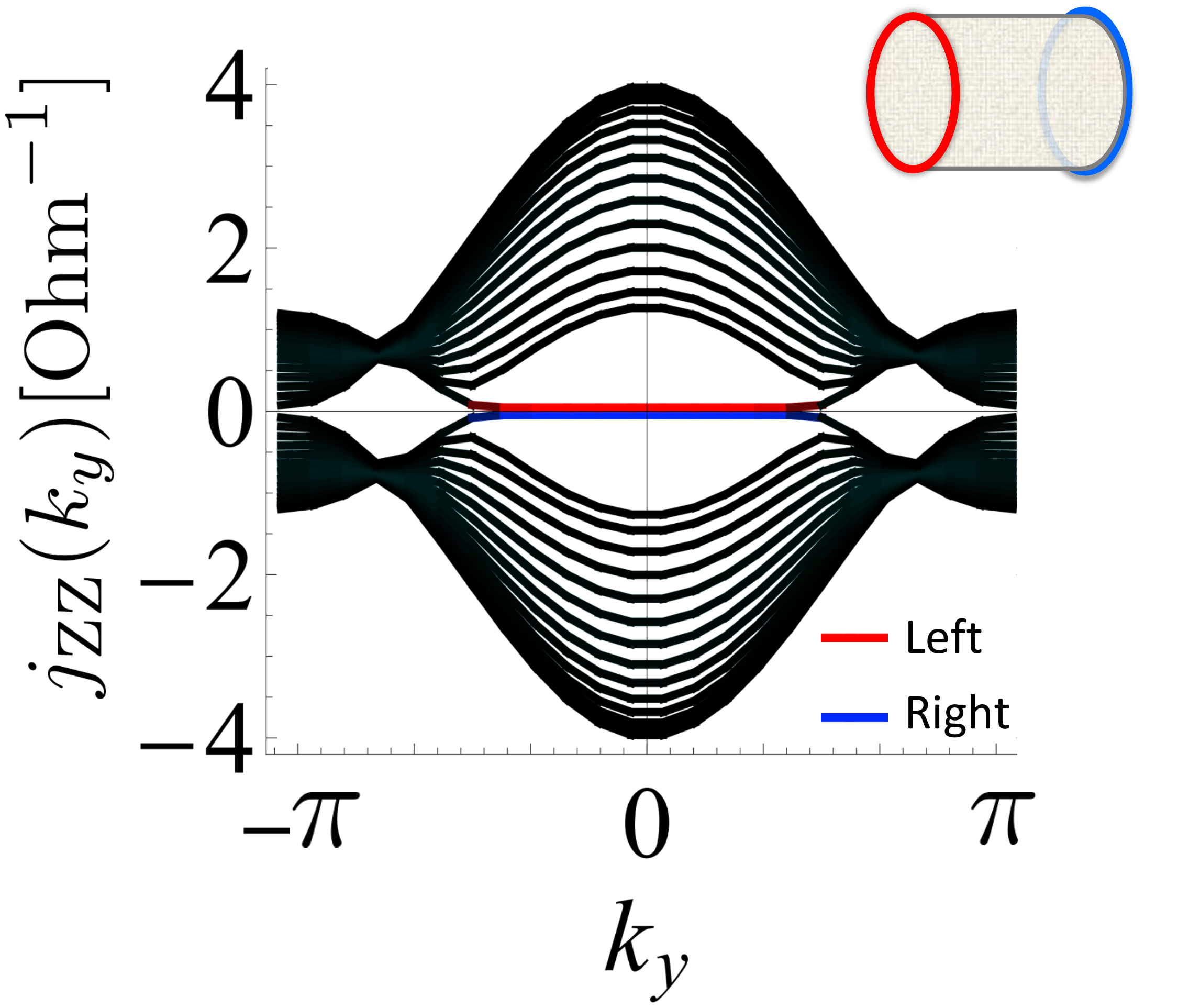}}
\subfloat[]{\includegraphics[width=0.54\linewidth]{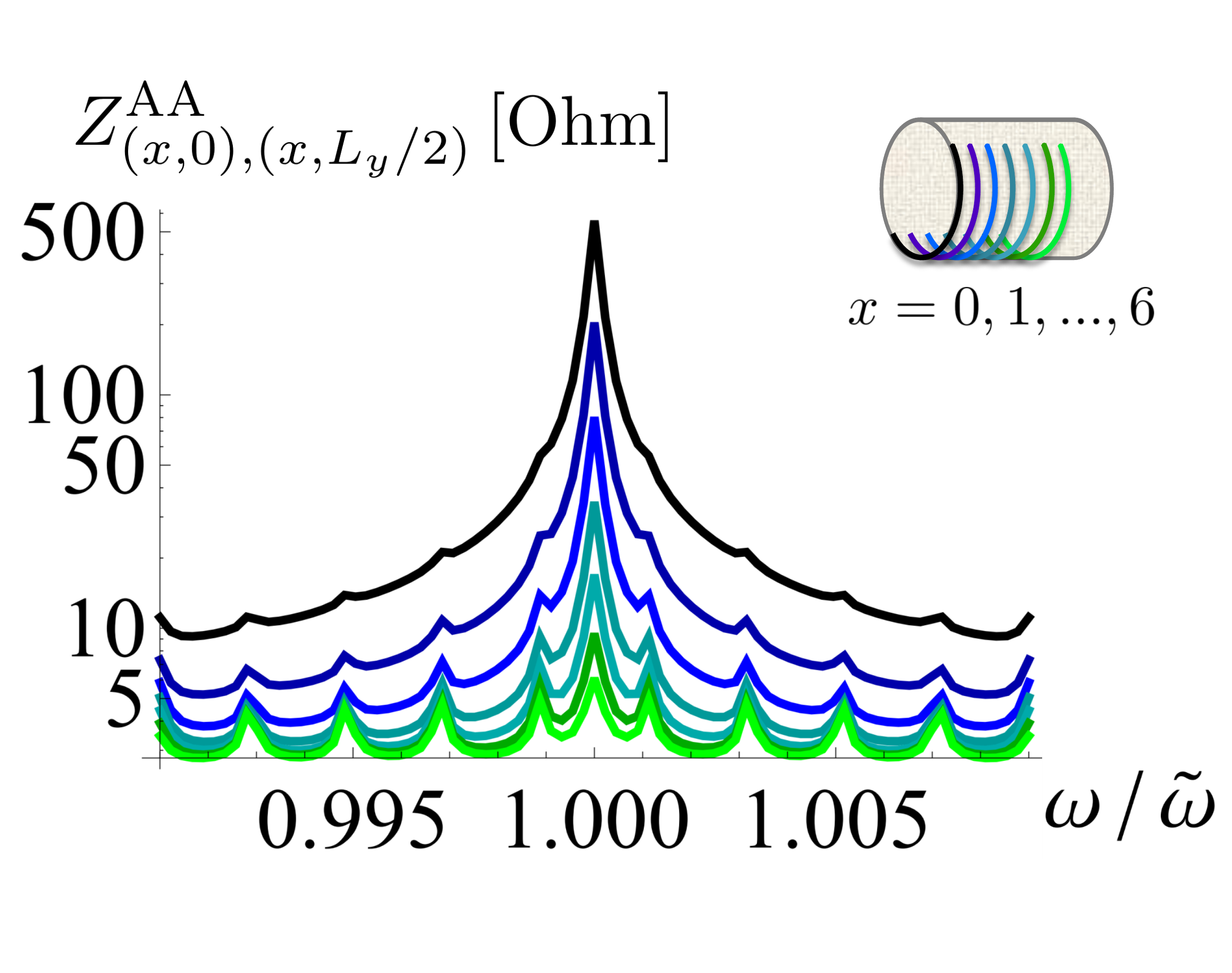}}\\
\subfloat[]{\includegraphics[width=0.99\linewidth]{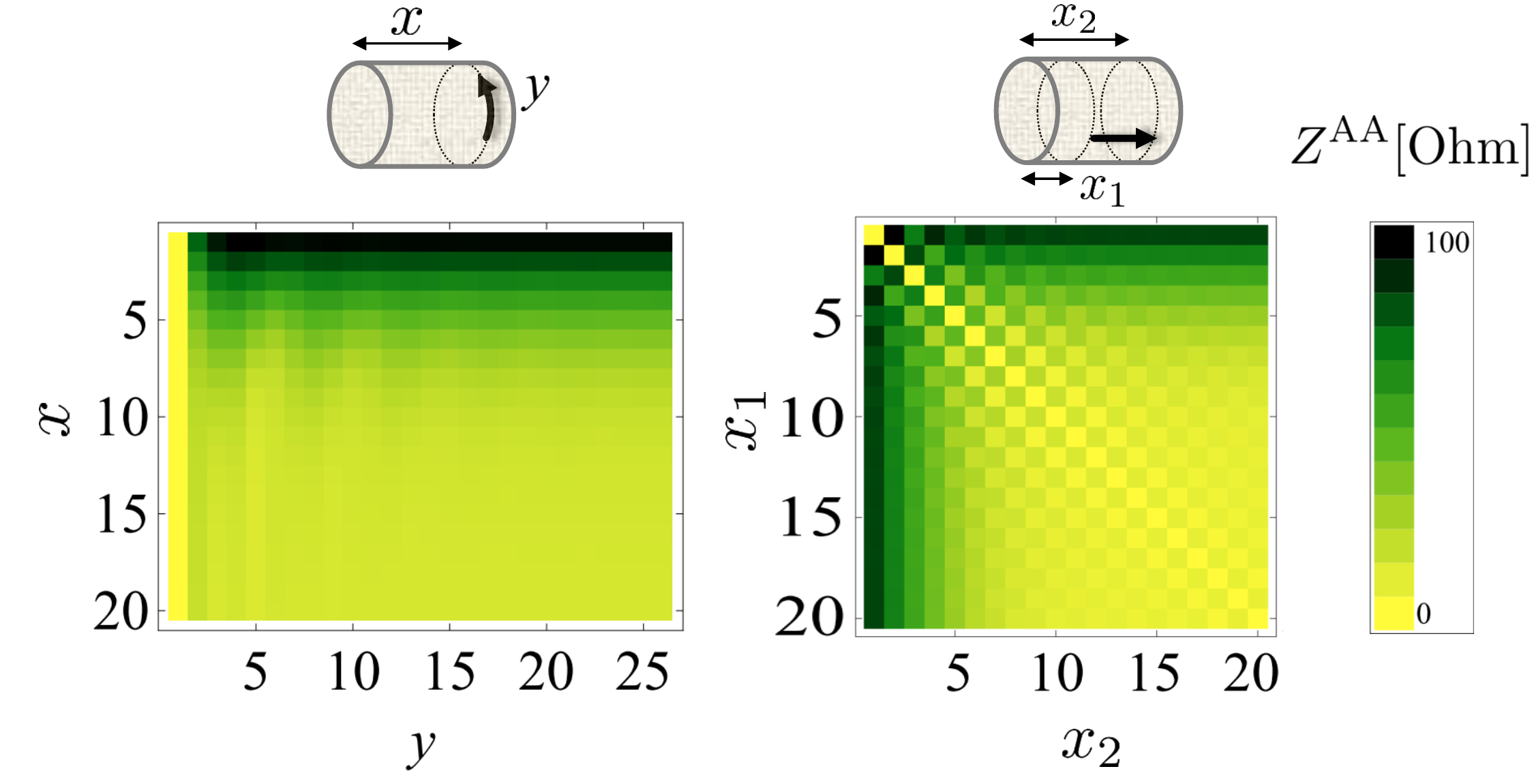}}
\end{minipage}
\caption{(a) Circuit diagram of the 2D zigzag topolectrical circuit of Eq.~\ref{J2}. Three types of capacitors connect the nodes in three different orientations, resulting in a two-site unit cell with sublattices denoted by A and B. The circuit is periodically connected in the $y$-direction with circumference $L_y$, and grounded on its left and right edges. (b) Bandstructure of $j_{ZZ}(k_y)$ with resonant frequency $\tilde\omega=10^3s^{-1}$, $L_y=12$, and capacitor parameters $(\alpha,\beta,\gamma)=(0.8,0.2,1) \text{mF}$. At both edges (red/blue), lines of degenerate modes with $j_{0,k_y} \approx 0$ span across the extensive range $|k_y|<\pi/2$. This leads 
to the TBR depicted in (c), which presents the impedance $Z^{\text{AA}}_{(x,0),(x,L_y/2)}$ between diametrically opposite nodes $x$ sites from the left edge ($x=0$). 
(d) presents the impedances $Z^{\text{AA}}_{(x_1,0),(x_2,0)}$ and $Z^{\text{AA}}_{(x,0),(x,y)}$ across intervals perpendicular and parallel to the edges. Note the rapid rise in impedance close to the edge.}
\label{circuit}
\end{figure}

A more targeted TBR response at the boundaries can be achieved in  higher-dimensional topolectrical circuits, where the increased admittance density of states (DOS) from an additional dimension makes it possible to spatially isolate the topolectrical resonance. The SSH circuit can be straightforwardly extended to represent a 2D band structure by 
adding a spatial modulation to the capacitances with inverse wavelength $k_y$ along a new direction, such that a phase transition at $t=1$ occurs at a certain range of $k_y$. This can be achieved, for instance, through the parametrization $C_1=\gamma+2\beta\cos k_y$, $C_2=\gamma+2\alpha\cos k_y$, bringing the grounded Laplacian to the form
\begin{eqnarray}
J_{\text{ZZ}}(\vec k)&=&i\omega\left(2(\gamma+\alpha+\beta)-\frac1{\omega^2L}\right)\mathbb{I}\notag\\
&& -i\omega(\gamma+2\beta\cos k_y+(\gamma+2\alpha\cos k_y)\cos k_x)\sigma_x\notag\\
&&-i\omega(\gamma+2\alpha\cos k_y)\sin k_x\,\sigma_y.
\label{J2}
\end{eqnarray}
In real space, this 2D circuit consists of a lattice network with two inequivalent nodes per unit cell, where unlike nodes are connected by capacitors of capacitances $\alpha,\beta$, or $\gamma$ depending on their relative orientations (Fig. \ref{circuit}). Each node is also connected to the ground by an inductor $L$. The two-site unit cell, along with the lattice connectivity and edge termination, provides a circuit analog of the zigzag (ZZ) edge of graphene~\cite{PhysRevB.54.17954}. This circuit network supports topological boundary modes inherited from its SSH predecessor. If we ground the capacitors on one/both of its edges perpendicular to the $x$-direction, but leave the circuit periodic in the $y$-direction by connecting the last capacitor with the first capacitor, a line of singly/doubly degenerate edge modes appear for $k_y$ satisfying $t<1$, i.e. $(\alpha-\beta)\cos k_y > 0$.  

\begin{figure*}
\begin{minipage}{\linewidth}
\subfloat[]{\includegraphics[width=0.17\linewidth]{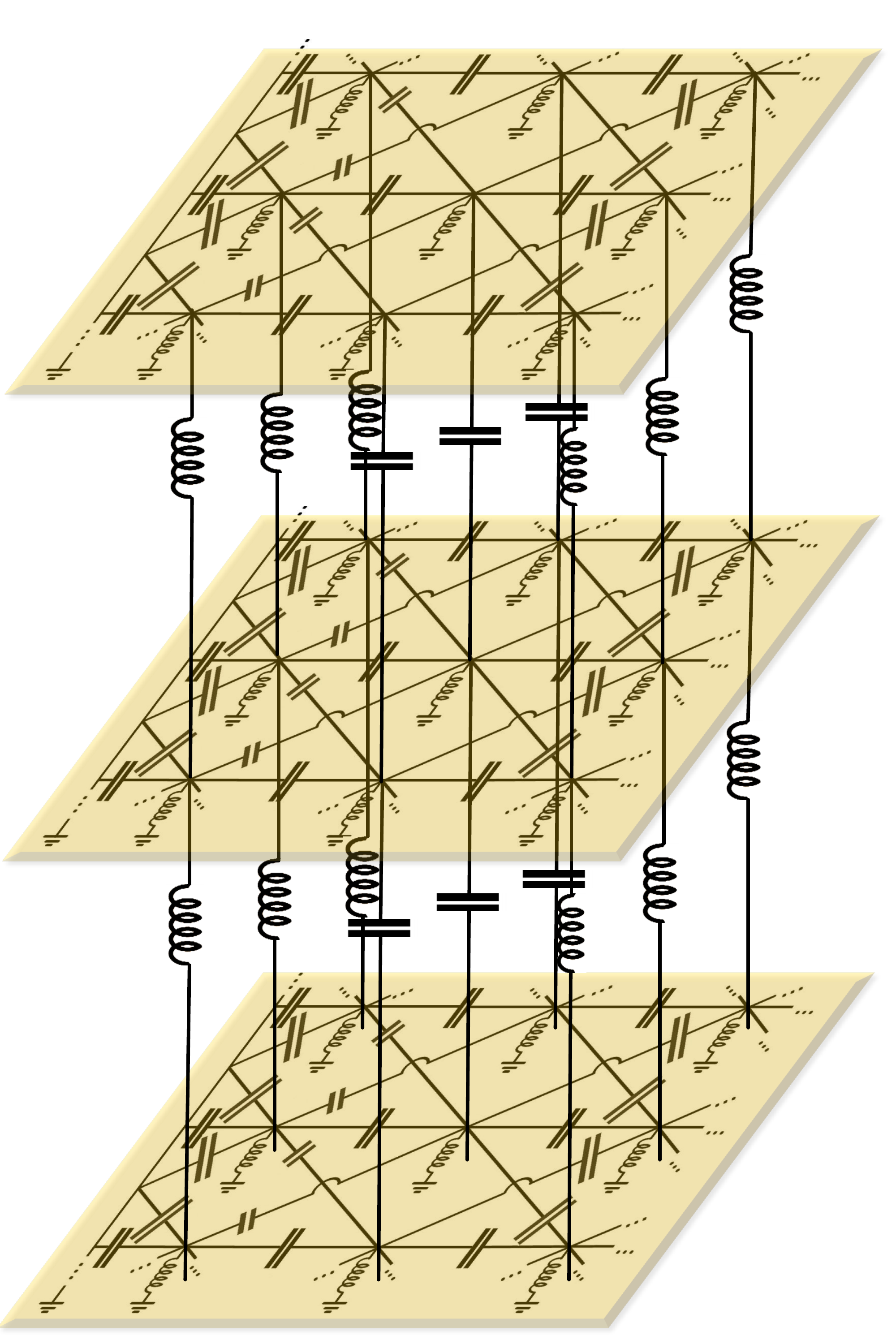}\label{topo_circuit_arc}}
\subfloat[]{\includegraphics[width=0.23\linewidth]{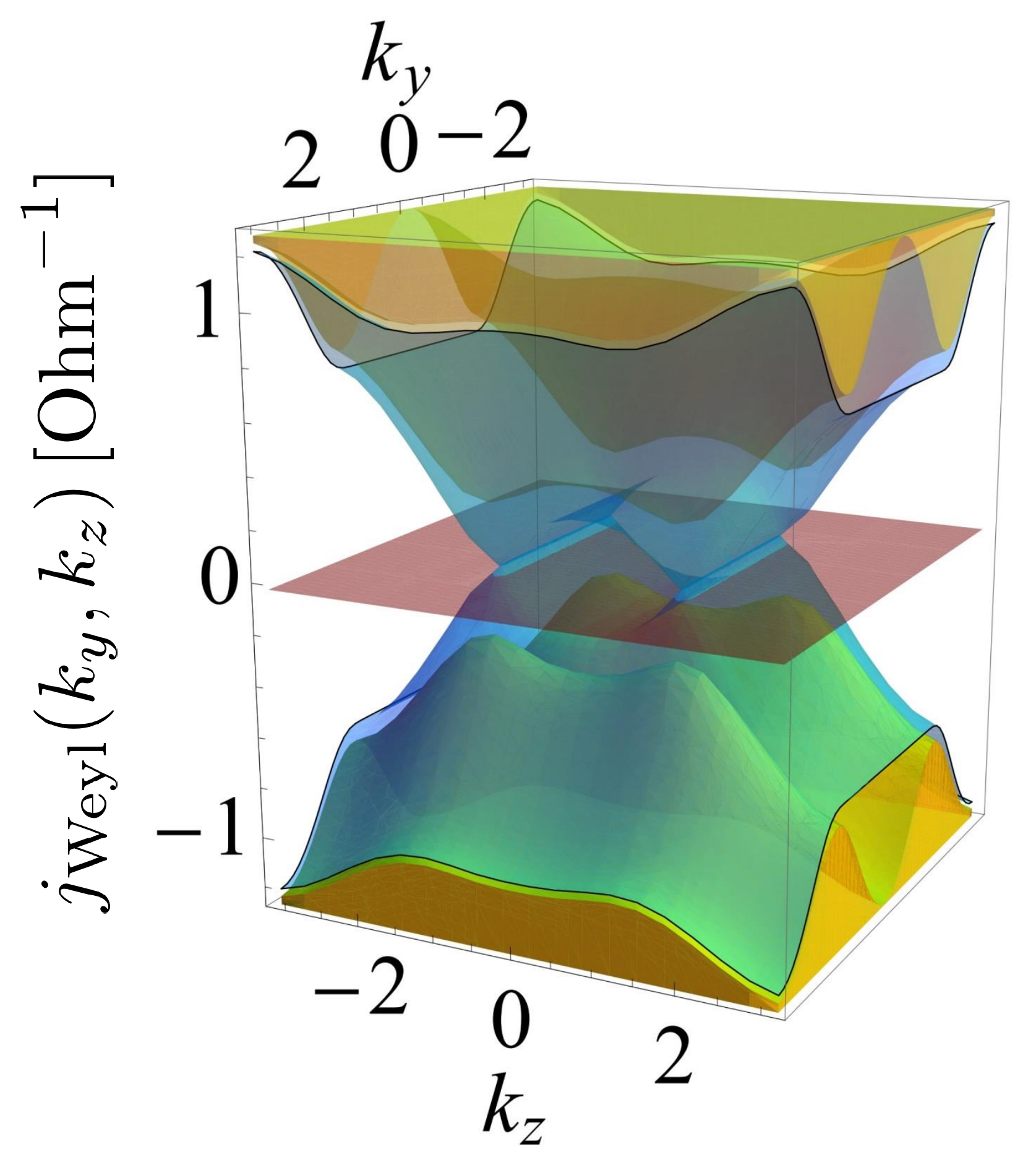}\label{topo_circuit_arc}}
\subfloat[]{\includegraphics[width=0.19\linewidth]{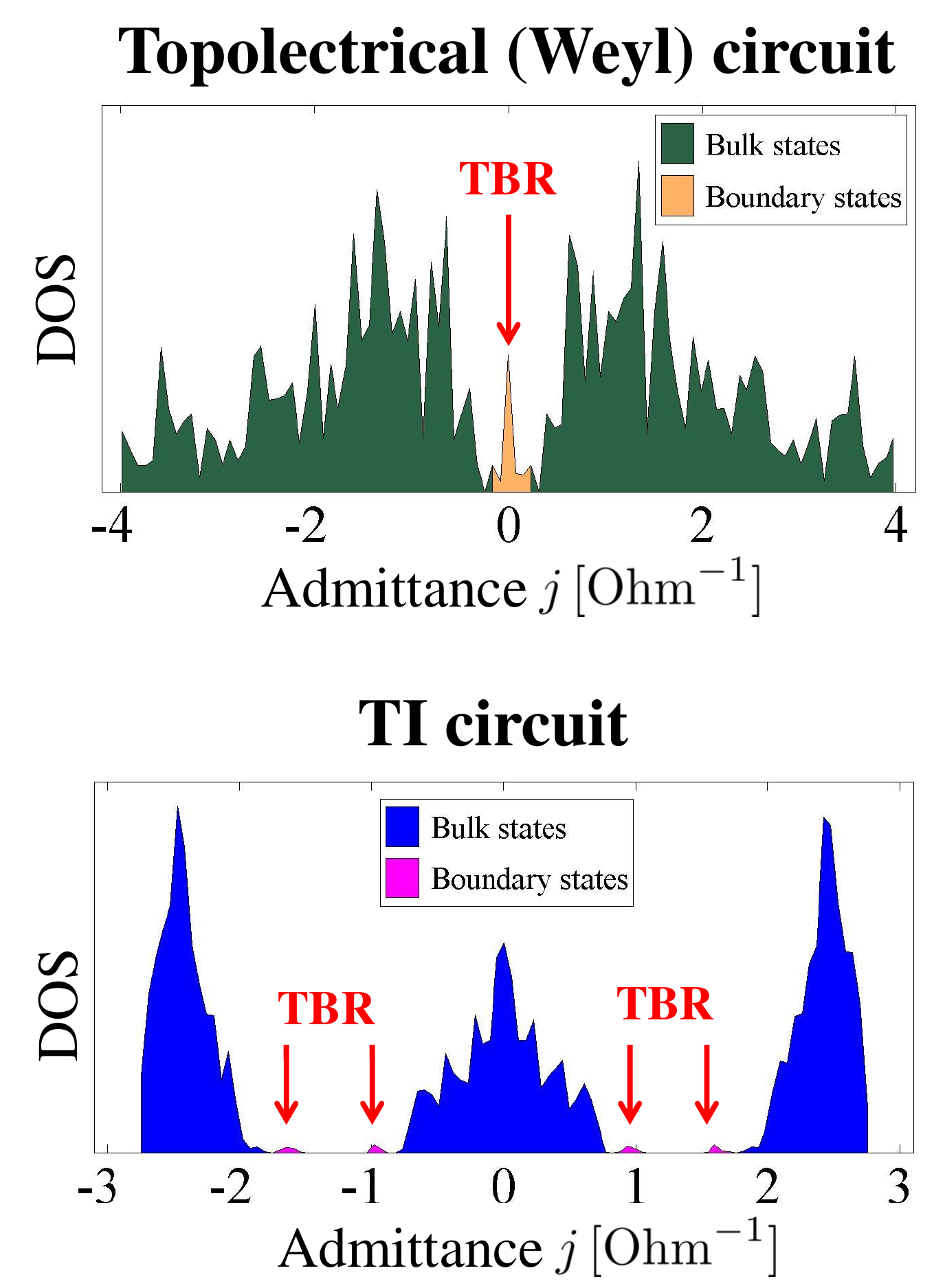}\label{topo_circuit_DOS}}
\subfloat[]{\includegraphics[width=0.17\linewidth]{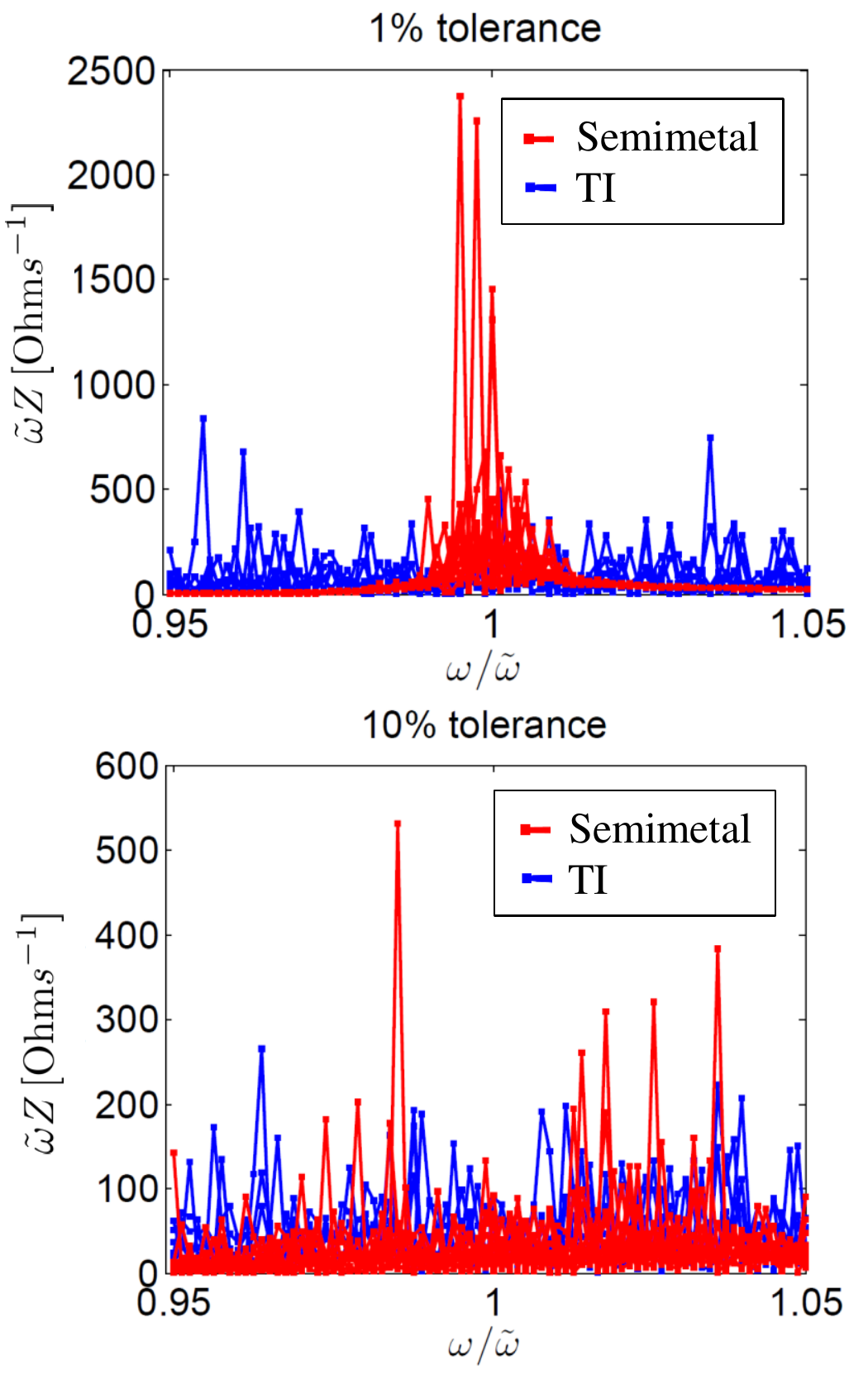}\label{topo_circuit_tolerance}}
\subfloat[]{\includegraphics[width=0.19\linewidth]{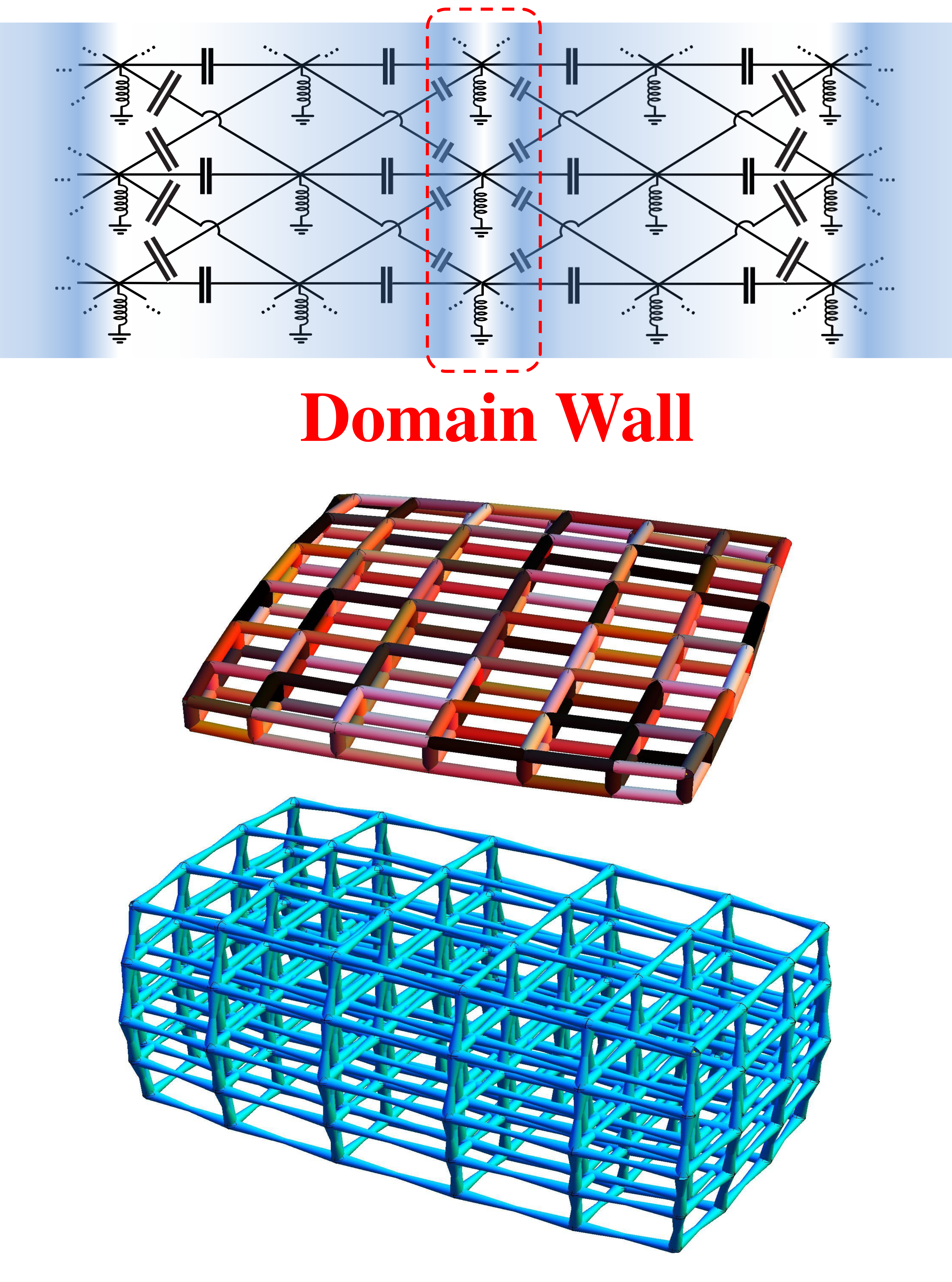}\label{topo_circuit_higherdim}}
\end{minipage}
\caption{(a) Circuit diagram of the Weyl semimetal topolectrical circuit (Eq. \ref{J3}). Upon appropriate termination in $x$ direction, it features resonant bands of modes analogous to Weyl semimetals, as depicted in (b) for parameters $\beta=0$, $\gamma=1$, $\alpha=\gamma_z=\lambda=1/2$. The Weyl circuit is grounded along the plane normal to the $x$-axis, just as in Fig.~\ref{circuit}. Its surface states (blue) separate from the bulk states (yellow), and intersect the $j_{\text{Weyl}}(k_y,k_z)=0$ plane (brown) along two straight Fermi arcs from $(k_y,k_z)=(-\pi/2,\pm \pi/3)$ to $(\pi/2,\pm \pi/3)$. (c) Comparison of the DOS of our semimetal topolectrical circuit with the grounded topological insulator circuit (see appendix), both with component nonuniformity tolerances of $1\%$. There are extensively more degenerate boundary modes contributing to the TBRs in the semimetal circuit, as compared to the dispersive TI edge modes traversing its 3 bands. (d) Impedance read-outs of a random ensemble of ten circuits of each type. 
Each capacitor $C$/inductor $L$ is also ascribed a resistive loss of $0.1\times\Delta|(i\tilde\omega C)^{-1}|$ and $0.1\times\Delta|i\tilde\omega L|$, where $\Delta C/C$ or $\Delta L/L$ are $1\%$ or $10\%$. The semimetal exhibits TBRs that are reasonably robust against disorder. This does not hold for the topological insulator circuit, whose resonances are mainly due to non-universal bulk modes. 
(e) Generalizations of the topolectrical circuits include domain wall states (Top), circuits with aperiodically modulated elements simulating topological quasicrystals (Middle) and circuits on hyperlattices (shown here with $6\times 4\times 4\times 3$ unit cells) hosting higher-dimensional states (Bottom). }
\label{higherdim}
\end{figure*}


When the AC frequency $\omega$ is tuned to the particular resonant frequency $\tilde\omega=\frac{1}{\sqrt{2L(\alpha+\beta+\gamma)}}$, these edge modes correspond to an extensive line of vanishing eigenvalues $j_0$ that dramatically enhance the circuit impedance at the edge. This can be physically explained in terms of edge resonances involving isolated triplets of simultaneously ``dimerized'' capacitors sharing oscillating charges, reminiscent of those in the SSH topolectrical circuit. When $\alpha>\beta$ and $|k_y|<\pi/2$, the capacitors in horizontally adjacent unit cells collectively ``dimerize'' by harboring strongly oscillating charges. Reversing the former condition to $\alpha<\beta$ makes the dimerization incompatible with the edge grounding, while breaking the condition $|k_y|<\pi/2$ also inhibits these oscillations by reversing the relative polarity of adjacent capacitors. Since this dimerization ultimately relies on sublattice symmetry, the TBR only appears when the edges respect sublattice symmetry, such as in the zigzag case. 
A realistic implementation is illustrated in Fig. \ref{circuit}, where a fixed realistic serial resistance is attached to each grounding wire. $Z^{ss'}_{{\bf x}_1, {\bf x}_2}$ denotes the impedance between the $s$th node
of unit cell ${\bf x}_1=(x_1,y_1)$ and the $s'$th node of unit cell ${\bf x}_2=(x_2,y_2)$, $s,s' \in \{\text{A,B}\}$, the left edge being located at $x=0$ (cf. appendices for detailed calculations). 
Topologically, the circuit is a cylinder periodic in $y$-direction, with a circumference of $L_y$ rows. As plotted in Fig.~\ref{circuit}d for A-type nodes, the impedances in both $x$ and $y$ directions ($Z^{\text{AA}}_{(x_1,0),(x_2,0)}$ and $Z^{\text{AA}}_{(x,0),(x,y)}$)  are greatly enhanced only near the edge, in contrast to the previous 1D SSH circuit. This enhancement is apparent even for short intervals, as reflected by the rapid rise of impedance $Z^{AA}_{(x,0),(x,y)}$ at small $y$. 
For circuits representing higher dimensional band structures, a deeper consequence of Eq. \ref{Z3} is that the TBR depends on the scale set by the maximal imaginary gap~\cite{lee2016band}, even if the boundary modes themselves are gapless and algebraically decaying. This conundrum is addressed in the appendix.  



The zigzag topolectrical circuit, which contains a line of zero eigenvalues when driven at resonant frequency, 
can be obtained as a slice of a parent 3D lattice of RLC elements with ``Fermi arcs'' in its bandstructure. One example is the circuit given, at resonance (see appendices), by
\begin{eqnarray}
J_{\text{Weyl}}(\vec k)|_{\omega=\tilde\omega}
&=& -i\tilde\omega(\gamma+2\beta \cos k_y+(\gamma+2\alpha \cos k_y)\cos k_x)\sigma_x\notag\\
&&-i\tilde\omega(\gamma+2\alpha\cos k_y)\sin k_x\,\sigma_y\notag\\
&&+2i\tilde\omega \gamma_z(1-\cos k_z -\lambda)\sigma_z,
\label{J3}
\end{eqnarray}
with capacitances, inductances, and resonant frequency satisfying $\tilde\omega^{-2}=2\text{L}(\alpha+\beta+\gamma)=\text{L}_z\gamma_z$. This circuit consists of layers of the zigzag topolectrical circuit (Fig.~\ref{circuit}a) connected by capacitors (inductors) of strengths $\gamma_z$($L_z$) above A (B) sublattice sites. Each A (B) site is additionally grounded by an inductor (capacitor) of strength $\frac{1}{2}\lambda^{-1}L_z$ ($2\lambda \gamma_z$). This circuit gives rise to TBRs that bear close similarity to Fermi arcs at zero Fermi energy as found in Weyl semimetals (Fig.~\ref{circuit}b). 
$J_{\text{Weyl}}(\vec k)|_{\omega=\tilde\omega}$ exhibits four Weyl points at $\vec k = (\pi, \pm \pi/2, \cos^{-1}(1-\lambda))$, which are connected by "Fermi arcs" along both branches of $ (\pi, k_y, \cos^{-1}(1-\lambda))$, $|k_y|<\pi/2$ (Fig.~\ref{topo_circuit_arc}). Along these Fermi arcs, we recover the line nodes of the zigzag topolectrical circuit, where the massive degeneracy is protected by sublattice symmetry. 


Topolectrical circuits can be realized using basic laboratory equipment. For a realistic implementation, however, as we have also seen for our experimental implementation of the SSH circuit, one has to take into account the non-uniformity of RLC components, as well as capacitive and resistive losses. We find that topolectrical circuits, which are analogous to topological semimetals, are markly superior to e.g. topological insulator circuits in this respect, as shown in Fig.~\ref{higherdim}c,d. There, we compare the impedance read-out of our topolectrical circuits containing extensive mode degeneracy (Eqs.~\ref{J2} and~\ref{J3}) with that of a topological insulator circuit~\cite{ningyuan2015time,PhysRevLett.114.173902}, which we adjusted by grounding it with inductors such that both systems can be accurately tuned for possible TBRs through the AC frequency (cf. appendices). Due to the extensively large boundary DOS which is broadened but not destroyed by disorder (Fig.~\ref{topo_circuit_DOS}), sharply defined topolectrical resonances exist e.g. in the Weyl or zigzag circuit, even with $10\%$ error tolerance in each circuit element. Furthermore, due to extensivity, the resonances remain pronounced even when disorder shifts them slightly away from $\tilde\omega$, i.e. the resonant frequency at zero disorder. By contrast, the impedance resonance peaks of the topological insulator circuit are neither as pronounced nor immune to disorder, since they are mostly due to the bulk modes. Although their boundary modes are likewise topologically protected, they exist at isolated momenta at any given frequency, and thus have limited contribution to the impedance read-out. We find, however, that the voltage eigenstate profile $\psi_0(n)$, as we have measured it for the SSH midgap mode, is still accessible, and thus is one of the most sensible quantities to measure for topological insulator circuits.


From a broader point of view on 3D topolectrical circuits, the requisite Fermi arcs for TBRs can occur in the presence of more exotic symmetries, e.g. certain non-symmorphic symmetries appearing in known Weyl semimetals or photonic crystals~\cite{wang2016hourglass}.
Many of these symmetries, and hence their accompanying topological phases, can be conveniently realized in electrical circuits, whose network structure is free from physical limitations imposed by the shape of ionic orbitals. Topolectrical circuits are likewise not restricted by intrinsic lengths scales from quantum mechanics, and can be constructed at macroscopic sizes with connections across arbitrarily distant nodes. In particular, topolectrical circuits can be used to simulate higher-dimensional topological phases without involving synthetic dimensions, since each node can be connected to other nodes along more than three axes (Fig.\ref{topo_circuit_higherdim}). The ability to go beyond a regular periodic structure also allows for an accessible study of topological phases on aperiodic networks~\cite{kraus2012topological} (Fig.\ref{topo_circuit_higherdim}) or hyperbolic lattices of arbitrary complexity~\cite{gu2016holographic,lee2016generalized}. Circuit elements such as capacitors can also be mechanically manipulated to break time-reversal symmetry, and induce novel non-equilibrium (Floquet) Chern phases~\cite{PhysRevB.79.081406}. In general, the TBR can also be designed to appear not at the physical boundary of the circuit, but at domain walls along an arbitrary trajectory through the circuit (Fig.\ref{topo_circuit_higherdim}). For the zigzag circuit, such a formulation would bear strong similarity to flux lattice domain walls~\cite{Sessi1269}. This domain wall design would certainly be of interest not just for topolectrical circuits, but also for mechanical systems.

\begin{acknowledgments}
We thank D. A. Abanin, H.-P. B\"uchler,  J. Chalker, Y.D. Chong, J. Gong, V. Manucharyan, R. Moessner, and  B. Yang for discussions. We acknowledge support by DFG-SFB 1170 TOCOTRONICS (project A07 and B04), by   ERC-StG-Thomale-336012-TOPOLECTRICS,  by ERC-AG-3-TOP, and by ERC-AG-4-TOPS.
\end{acknowledgments}

\section*{Author contributions}

CHL and RT initiated the project and contributed the theoretical analysis. SI, CB, FB, JB, LM, and TK performed the experiment on the SSH circuit and provided overall guidance in terms of experimental implementation. CHL, TK, and RT contributed to writing the manuscript.  

\section*{Competing financial interests}

The authors have no competing financial interests.

\bibliographystyle{naturemag}
\bibliography{bibliography_new,references}
\newpage
\begin{widetext}
\begin{center}
\textbf{\large Method Section (Appendices) for ``Topolectrical Circuits" }
\end{center}
\appendix
\section{Impedance formulae for 2-component periodic circuits}
It is instructive to explicitly evaluate Eq.~\ref{Z3} for general 2-node unit cell circuits for periodic boundary conditions (i.e. without grounded terminations). We start from
\begin{equation}
Z^{s,s'}_{\vec x}=\sum_{\vec k,m}\frac{|\phi_{m}(\vec k,s)-\phi_{m}(\vec k,s')e^{i\vec k\cdot \vec x}|^2}{j_{\vec k,m}},
\end{equation}
where $\vec k$ denotes momentum, and $\phi_m$ are the eigenvectors of $J(\vec k)=d_0(\vec k)+\vec d(\vec k)\cdot \vec{\sigma}$, with $\vec\sigma$ being the vector of Pauli matrices. In closed form, the impedances between nodes of sublattices A, B separated by $\vec x$ unit cells are given by (with $\hat d = \vec d / |\vec d|$)
\begin{align} 
Z^{\text{AA}}_{\vec x}&=\sum_{\vec k,\pm}\frac{1\pm \hat d_3}{d_0\pm |d|}(1-\cos\vec k\cdot \vec x), \\ 
 Z^{\text{BB}}_{\vec x}&=\sum_{\vec k,\pm}\frac{1\mp \hat d_3}{d_0\pm |d|}(1-\cos\vec k\cdot \vec x), \\
 Z^{\text{AB}}_{\vec x}&=\sum_{\vec k,\pm}\frac{1\pm\sqrt{1-\hat d_3^2}\cos\left(\vec k\cdot \vec x + \tan^{-1}\frac{d_2}{d_1}\right)}{d_0\pm |d|}, \\
 Z^{\text{BA}}_{\vec x}&=\sum_{\vec k,\pm}\frac{1\pm\sqrt{1-\hat d_3^2}\cos\left(\vec k\cdot \vec x - \tan^{-1}\frac{d_2}{d_1}\right)}{d_0\pm |d|} .
\end{align}

\section{Circuit Green's function}
A physical interpretation of $G=J^{-1}$ as the inverse of a graph Laplacian is readily obtained. Write $J=D+W-C$, where $[D]_{ab}=\delta_{ab}\sum_{c}\text{C}_{ac}$ is the diagonal matrix of the conductances emanating from each node towards other nodes, $[W]_{ab}=\delta_{ab}w_a$ the conductance of each node towards the ground, and $C$ the adjacency matrix of the conductances. Then
\begin{eqnarray}
G=\frac1{D+W-C}=(D+W)^{-1}\sum_{n=0}^\infty ((D+W)^{-1}C)^{n}
\end{eqnarray}
i.e. $G_{ab}$ is the number of paths of any length from node $a$ to $b$, each weighted by the conductance ratio (i.e. transition probability) $[(D+W)^{-1}C]_{kl}=\text{C}_{kl}/(w_k+\sum_{l'} \text{C}_{kl'})$ between each pair of nodes $k,l$ along the path. In other words, $G$ keeps track of the fraction of the current that will flow between two nodes, assuming that it spreads out at each node it passes by.

\section{SSH circuit}
\subsection{Ideal analytical solution}
The simplest topolectrical circuit can be written out in explicit but still compact detail. From Kirchhoff's law, we can write the grounded Laplacian as 
\begin{eqnarray}
J_{\text{SSH}}&=&\left(\begin{array}{cccccc}
i\omega(C_1+C_2)+\frac1{i\omega L} & -i\omega C_1 & 0 & 0 & 0  & ... \\
-i\omega C_1 & i\omega(C_1+C_2) +\frac1{i\omega L}& -i\omega C_2 & 0 & 0  & ... \\
0 & -i\omega C_2 & i\omega(C_1+C_2)+\frac1{i\omega L} & -i\omega C_1 & 0  & ... \\
0 & 0 & -i\omega C_1 & i\omega(C_1+C_2)+\frac1{i\omega L} & -i\omega C_2  &  ... \\
0 & 0 & 0 & -i\omega C_2 & i\omega(C_1+C_2)+\frac1{i\omega L}   &  ... \\
\vdots & \vdots & \vdots & \vdots & \vdots    &  \ddots \\
\end{array}\right)\notag\\
&=&i\omega C_2\left(\begin{array}{cccccc}
(1+t)\left(1-\frac{\tilde \omega^2}{\omega^2}\right) & -t & 0 & 0 & 0  & ... \\
-t & (1+t)\left(1-\frac{\tilde\omega^2}{\omega^2}\right)& -1 & 0 & 0  & ... \\
0 & -1 & (1+t)\left(1-\frac{\tilde\omega^2}{\omega^2}\right)& -t & 0  & ... \\
0 & 0 & -t & (1+t)\left(1-\frac{\tilde\omega^2}{\omega^2}\right) & -1  &  ... \\
0 & 0 & 0 & -1 & (1+t)\left(1-\frac{\tilde\omega^2}{\omega^2}\right)  &  ... \\
\vdots & \vdots & \vdots & \vdots & \vdots    &  \ddots \\
\end{array}\right),
\label{Jij}
\end{eqnarray}
with 
the last diagonal entry containing $C_1$, $C_2$ or both depending on which type of capacitor (or both) is connected to the rightmost node. When $\omega$ is set to the resonant frequency $\tilde\omega =\frac1{\sqrt{L(C_1+C_2)}}$, the diagonal terms proportional to the identity disappear, and $J$ possesses an exact expression for its inverse which is given by
\begin{eqnarray}
&&G=\frac1{i\omega (-t)^{n}C_2 }
\left(
\begin{array}{ccccccc}
1  & (-t)^{n-1} & -t & (-t)^{n-2} & (-t)^2 & (-t)^{n-3} & ...\\
(-t)^{n-1}  & 0 & 0 & 0 & 0 & 0 & ...\\
-t  & 0 & (-t)^2 & (-t)^{n-1} & (-t)^3 & (-t)^{n-2} & ...\\
(-t)^{n-2}  & 0&(-t)^{n-1}  & 0  & 0 & 0 & ...\\
 (-t)^2 & 0 & (-t)^3 & 0 & (-t)^4 & (-t)^{n-1} & ...\\
 (-t)^{n-3}  & 0& (-t)^{n-2}  & 0&(-t)^{n-1}   & 0 & ...\\
\vdots & \vdots & \vdots & \vdots & \vdots  &\vdots  &  \ddots \\
\end{array}
\right)
\end{eqnarray}
For $C_1/C_2=t<1$, there exists a boundary mode near $(1+t)\left(1-\frac{\tilde\omega^2}{\omega^2}\right)$, the middle of the bulk spectral gap of $J_{\text{SSH}}$. Its eigenvalue $j_0$ can be obtained from the characteristic polynomial of $J_{\text{SSH}}$. At resonant frequency $\omega=\tilde\omega$, $j_0$ is very close to zero, and we can neglect all but the linear term of the characteristic polynomial to obtain 
\begin{equation}
j_0\approx i\tilde\omega\frac{(-t)^N(1-t^2)}{1-t^{2\left\lfloor N/2\right\rfloor}},
\label{j0resonance}
\end{equation}
which exponentially decays with $N$, the number of capacitors in the SSH circuit. From Eq.~\ref{Z2} and $\psi_0\propto (1,0,-t,0,t^2,...)$, the impedance between nodes $1$ and $2x-1$ (or $2x$) is thus given by 
\begin{align}
Z_{1,2x-1}^{\text{SSH}}&=\frac{|\psi_0(1)-\psi_0(2x-1)|^2}{j_0}=\frac1{i\tilde\omega C_2}\frac{\left(1-(-t)^{x-1}\right)^2}{(-t)^N},\\
Z_{1,2x}^{\text{SSH}}&=\frac{|\psi_0(1)-\psi_0(2x)|^2}{j_0}=\frac1{i\tilde\omega C_2}\left((-t)^{-N}-2(-t)^{-x}\right),
\label{Z_1}
\end{align}
both of which are plotted in Fig.~\ref{circuit1D}. Note that had $C_2$ been instead greater than $C_1$, $\psi_{0}(x)$ will not be have  been able to exist as an eigenmode due to incompatible boundary conditions. However, if the right end of the circuit is also connected to a grounded capacitor $C_2$, there can be another mirror-reflected, but otherwise identical, boundary mode localized on the right end.

To find the momentum space representation of the grounded Laplacian, we impose  periodic boundary conditions and Fourier transform Eq. \ref{Jij} to obtain
\begin{eqnarray}
J_{\text{SSH}}(k_x)&=&i\omega\left(C_1+C_2-\frac1{\omega^2L}\right)\mathbb{I}\notag\\
&& -i\omega\left[(C_1+C_2\cos k_x)\sigma_x+C_2\sin k_x\sigma_y\right],
\label{J1a}
\end{eqnarray}
as also presented in the main text. 

From Eq.~\ref{J1a}, it can be shown that for $t<1$, $J$ possesses gapped translation-invariant eigenmodes with eigenvalues $j_{k_x,\pm}=C_1+C_2-\frac1{\omega^2L}\pm \sqrt{C_1^2+C_2^2+2C_1C_2\cos k_x}$. As follows from Eq. \ref{j0resonance}, there is also a midgap boundary mode with the eigenvalue 
\begin{equation}
j_0\approx C_1+C_2-\frac1{\omega^2L}+(-t)^{N},
\end{equation}
which is not small when away from resonance. The SSH circuit has the special property that the decay length $\xi=\log \frac{C_2}{C_1}$ of its boundary mode coincides exactly with the imaginary gap~\cite{he2001exponential,lee2016band}, which is the imaginary part of $k_x$ necessary for closing the gap:
\begin{equation}
\sqrt{C_1^2+C_2^2+2C_1C_2\cos k_x}=0\; \; \Rightarrow e^{ik_x}=t\;\text{or}\; t^{-1}.
\end{equation}



On hindsight, the TBR of the SSH circuit has an elementary interpretation: Due to the special potential profile of the boundary mode (Fig. \ref{circuit1D}), a driving input voltage $V_0$ and current $I_0$ is connected to two capacitors and one inductor, all of which are at zero potential (grounded) at the other end. From the special decaying potential profile, the potential towards the far right (call it node $b$) must have vanishing potential. By Kirchhoff's law, $I_0=\left(i\tilde\omega(C_1+C_2)-\frac1{i\tilde\omega L}\right)V_0\rightarrow 0$. Hence follows the impedance $Z_{1b}=\frac{V_0-V_b}{I_0}\approx \frac{V_0}{I_0} \rightarrow \infty$.

\subsection{Sublattice symmetry ($Z_2$) and SSH winding number}
Eq.~\ref{J1a} is a map from $S^1$ to $S^1$, and is characterized by an integer winding number
\begin{eqnarray}
N_{1D}&=&\frac1{2\pi}\oint\vec A\cdot dk\notag\\
&=&-\frac{i}{2\pi}\oint\phi^\dagger\nabla\phi~ dk_x\notag\\
&=&\int_{k_x=0}^{k_x=2\pi} d\left[\tan^{-1}\left(\frac{\sin k_x}{C_1/C_2+\cos k_x}\right)\right]\notag\\
&=&\theta(C_2-C_1).
\label{N1}
\end{eqnarray}
The second step of Eq. \ref{N1} relies on the absence of a $\sigma_z$ term in $J_{\text{SSH}}(k_x)$, which is enforced by the sublattice symmetry of the circuit (every node looks the same up to a left-right reflection). This winding woul not be well-defined if, for instance, different inductors were connected to different nodes.
This winding number $N_{1D}$ is related to the Chern number in the following sense. Consider a 2D extension to $J_{\text{SSH}}$ (Eq. \ref{J1a}). The contribution to $N_{2D}$ from a reciprocal space region $R$ is given by
\begin{eqnarray}
N_{2D}&=&\frac1{2\pi}\int_R \nabla\times \vec A~ d^2k\notag\\
&=& \frac1{2\pi}\oint_{\partial R} \vec A\cdot dk\notag\\
&=& \frac1{4\pi}\int_R \vec d\cdot (\partial_x\vec d\times \partial_y \vec d)~d^2k
\end{eqnarray}
where $\nabla\times \vec A$ is the Berry flux. Now suppose there is sublattice symmetry, i.e. that there is no $\sigma_3$ term so that $\vec d \perp \hat e_3$. Then the second line reduces to a line integral along the equator of the bloch sphere, which is mathematically known as $S^1$. Consequently, we can define a \emph{one-dimensional} topological invariant of the winding of the mapping from the 1D torus $\partial R$ to the equator $S^1\rightarrow S^1$, as we did above. This invariant needs the protection of sublattice symmetry; upon breaking it by adding a small $\sigma_3$ term, the $\vec d$ vector will not be confined to the equator, and a second homotopy invariant instead of a first homotopy invariant is required.

\subsection{Realistic circuit calculation}
With series resistance $R$ on the inductor, which is the most relevant serial resistance to consider, we have the impedance of each inductor replaced by $i\omega L \rightarrow i\omega L + R$. The grounded Laplacian is hence replaced by 
\begin{equation}
J_{SSH}=i\omega\left[\left(C_1+C_2-\frac1{\omega^2(L+R/(i\omega))}\right)\mathbb{I} -(C_1+C_2\cos k_x)\sigma_x-C_2\sin k_x\sigma_y\right].
\label{SSHR}
\end{equation}
The TBR resonance peak occurs when the magnitude of the identity matrix term is as small as possible, i.e at the minimal value of 
\begin{eqnarray}
Z^{-1}_{Res}&=&\left | i\omega(C_1+C_2)+\frac1{i\omega L +R}\right|\notag\\
&=& \frac{|(1-\omega^2 L (C_1+C_2))+i\omega R (C_1+C_2)|}{\sqrt{R^2+\omega^2L^2}}\notag\\
&=& \frac{\sqrt{(1-\omega^2 L (C_1+C_2))^2+\omega^2 R^2 (C_1+C_2)^2}}{\sqrt{R^2+\omega^2L^2}},\notag
\end{eqnarray}
which occurs at $\omega^2=\tilde\omega^2 = \frac1{L(C_1+C_2)}\sqrt{1+\frac{2(C_1+C_2)R^2}{L}}-\frac{R^2}{L^2}=\frac1{L(C_1+C_2)}(\sqrt{1+2\alpha}-\alpha)\approx \frac1{L(C_1+C_2)}\left(1-\frac1{8}\alpha^2\right)$ for small $\alpha=\frac{R^2(C_1+C_2)}{L}$, as can be checked via finding the extremum of the above. For a resonance to occur at a nonzero real frequency, we will need $\alpha < 1+\sqrt{2}\approx 2.41$, while an accessible resonance realistically requires $\alpha < 10^{-3}$ judging from our simulations. For $R=28\,\text{mOhm}$, $L=10^{-5}H$ and $C_1+C_2=1.22\times 10^{-7}F$ as given in our experimental setup, we have $\alpha \approx 10^{-5}$, which is sufficiently small for a clean and visible resonance. Circuit element non-uniformities hardly have any detrimental effect on the SSH signal, as we checked up to $20\%$ tolerance.

\subsection{Experimental implementation}

For the experimental implementation of the SSH-circuit a printed circuit board hosting ten unit cells was designed and fit with low serial resistance ($< 26 m\Omega$) inductors (Coilcraft MA5172-AE) and surface mount multilayer ceramic chip capacitors (Kemet 0805 / 1206), respectively. The circuit was fed by an arbitrary waveform generator (Agilent 33220A), the signals were picked up by lock-in amplifier (Zurich Instruments MFLI series).     

\section{Zigzag circuit}
As explained in the main text, the analysis of the zigzag topolectrical circuit is easily understood from the viewpoint of employing individual SSH circuits as building blocks (Fig. 3), which is why only few supplementary comments are needed. In real space, the edge mode consists of a superposition of various momenta $k_y$, each having the decay length of $\xi=(-\log t)^{-1}$, where $t=\frac{\gamma+2\beta\cos k_y}{\gamma+2\alpha\cos k_y}$. The profile of the real space mode is dominated by the slowest decaying $k_y$ mode, and decays algebraically whenever $t=1$ for some $k_y$. In our case, this occurs at $\cos k_y =0$, or everywhere if $\alpha=\beta$.
It now may appear that the capacitors of type $\gamma$ do not affect the qualitative decay of the edge modes. They, however, certainly affect the decay rate of the subdominant contributions, where $\cos k_y\neq 0$. Furthermore, they also affect the Laplacian bulk spectral gap which is given by $4i\omega\,\text{min}(|\alpha-\beta|,|\alpha-\gamma+\beta|)$, and hence affect the signal to noise ratio of the TBR. 

The strength of the TBR depends on the length scale set by the \emph{largest} imaginary gap, which here coincides with $-\log t$, even when the edge modes decay algebraically. To understand why, note that the TBR depends on the divergence of the impedance contributions from \emph{all} eigenmodes (Eq.~\ref{Z2}), especially the modes with \emph{smallest} $\text{min}(t,t^{-1})$. By contrast, the decay rate depends on the nature of the mode with the \emph{largest} $\text{min}(t,t^{-1})$; if the latter is unity, we have algebraic decay even though the TBR stems from the most strongly gapped momentum.

\section{Weyl circuit}
The Weyl circuit consists of layers of the zigzag topolectrical circuits (Fig. \ref{circuit}) connected by capacitors (inductors) of strengths $c_z$ ($L_z$) at A (B) sublattice sites (Fig. 4a). Each A (B) site is additionally grounded by an inductor (capacitor) of strength $\frac1{2}\lambda^{-1}L_z$ ($2\lambda \gamma_z$). Its grounded Laplacian is hence given by
\begin{eqnarray}
J_{\text{Weyl}}(\vec k)&=&i\omega\left(2(\gamma+\alpha+\beta)-\frac1{\omega^2L}+\left(\gamma_z-\frac1{\omega^2L_z}\right)(1-\cos k_z -\lambda)\right)\mathbb{I}\notag\\
&& -i\omega(\gamma+2\beta\cos k_y+(\gamma+2\alpha\cos k_y)\cos k_x)\sigma_x\notag\\
&&-i\omega(\gamma+2\alpha\cos k_y)\sin k_x\,\sigma_y\notag\\
&&+i\omega \left(\gamma_z+\frac1{\omega^2L_z}\right)(1-\cos k_z -\lambda)\sigma_z
\label{J3full}
\end{eqnarray}
with the resonant frequency $\tilde\omega$ given by $2L(\alpha+\beta+\gamma)=L_zc_z=\tilde\omega^{-2}$. Eq.~\ref{J3full} reduces to Eq. \ref{J3} at resonance.

\section{Robustness of the topological semimetal paradigm: semimetal circuit vs. topological insulator circuit}

To illustrate the significance of the semimetal paradigm in the light of the few existing works on electrical circuit realizations of topological phases, we present a detailed comparative analysis of the impedance read-out of our Weyl and zigzag circuits with those of Refs.~\onlinecite{ningyuan2015time,PhysRevLett.114.173902}, which realize a topological insulator phase through arrangements of circuit elements possessing appropriate internal symmetry. For a meaningful comparison, we slightly modified their topological insulator circuit by adding grounded capacitors $C$ to every node, so that the AC driving frequency indeed takes the role of the chemical potential in the topological insulator circuit as well.

In the construction of Ref.~\onlinecite{PhysRevLett.114.173902}, which generalizes that of Ref.~\onlinecite{ningyuan2015time} to arbitrarily large unit cells, the key idea is to realize a topologically nontrivial $Z_2$ phase protected by a geometric analog of the electronic antiunitary time-reversal operator. Although any RLC circuit must be time-reversal symmetric, it is possible to achieve a nontrivial $Z_2$ invariant by stacking together two copies of Hofstadter models with opposite magnetic fluxes, entangled in such a way that there is no need of realizing two spatially separated Chern circuits. By connecting unit cells with internal cyclic permutation symmetry with inductors that implement cyclic permutation operations, the electronic time reversal operator is mapped to a combination of ordinary, i.e. non-projective, time-reversal operations and cyclic permutations.

In our notation, the simplest topological insulator circuit, which contains 3 capacitors per magnetic unit cell (Eq. 3 of Ref.~\onlinecite{PhysRevLett.114.173902}), possesses the effective grounded Laplacian consisting of two copies ($\pm$) of
\begin{eqnarray}
J_{\text{TI}}(\vec k)&=&i\omega \left[L_1\left(
\begin{array}{ccc}
 0 & -1 & -e^{ik_x} \\
 -1 & 0 & -1 \\
 -e^{-ik_x} & -1 & 0
\end{array}
\right)+L_2\left(
\begin{array}{ccc}
 -2\cos k_y & 0 & 0 \\
 0 & -2\cos (k_y\pm 2\pi/3) & 0 \\
 0 & 0 & -2\cos (k_y\mp 2\pi/3)
\end{array} 
\right)+ \right. \nonumber \\
&&\left.  2(L_1+L_2)\left(1-\frac{\tilde\omega^2_{\text{TI}}}{\omega^2}\right)\mathbb{I}_{3\times 3}\right] \label{qsh},
\end{eqnarray}
where $\tilde\omega^2_{\text{TI}}=\frac1{2C(L_1+L_2)}$. Note that, opposite to our semimetal circuits, but in accordance to the convention in Ref.~\onlinecite{PhysRevLett.114.173902}, the ungrounded elements are the inductors, not capacitors. This duplicity yields no extra complication, as the simple relation $\omega \rightarrow \frac{\tilde\omega^2}{\omega}$ holds when the capacitors and inductors are interchanged.
Eq.~\ref{qsh} is a variant of (2 opposite copies of) the 3-band Hofstadter Hamiltonian, with each copy possessing 3 bulk bands connected by topological edge modes. The crux is that although these are bona fide topologically protected modes, they do not necessarily contribute significantly to the RLC resonances because they cross a given eigenvalue only at isolated points in momentum space. According to our semimetal paradigm, TBRs are characterized by boundary modes that are (i) extensively degenerate and (ii) spatially localized, with the former not being satisfied by the TI circuit edge mode(s).

To study the precise implications of the absence of extensive degeneracy in a topolectrical circuit, we consider ensembles of disordered circuits, i.e. circuits consisting of elements with nonuniform capacitances $C$ or inductances $L$. The nonuniformities are charactized by standard errors with tolerances (standard deviations) of $1\%$ or $10\%$. Additionally, we have included resistive losses proportional to $10\%$ of the non-uniformities of impedances due to disorder.
The results are depicted in Figs.~\ref{topo_circuit_DOS} and~\ref{topo_circuit_tolerance}. It becomes evident that the extensive, semimetal-like degeneracy of our Weyl circuit protects the RLC resonances much better than the protection from the single mode Kramer's degeneracy in the TI circuit. This observation identically holds for both the Weyl and zigzag topolectrical circuits.

\section{Transferring the topolectrical semimetal paradigm to other classical systems}

\subsection{Mechanical systems}
RLC circuits obey a linear 2nd order ODE, just like a mechanical system with springs, dampers and masses, and hence brings up the natural question what the mechanical system analogs of topolectrical circuits are. 
Topological mechanical systems have already been intensely studied in recent years, although their responses are not typically characterized by an impedance measurement. In a mechanical system with a single polarization direction (e.g. mechanical graphene~\cite{socolar2016mechanical}), the equation of motion likewise involves the Laplacian: $Lx=M\ddot{x}=-\omega^2 Mx$, i.e. 
\begin{equation}
J\vec x=(L-\omega^2M)\vec x.
\end{equation}
Since the mass matrix is diagonal, it is almost trivial to turn the above into an eigenvalue equation of $LM^{-1}$, with eigenvalues being $\omega^2$. Nonzero density of states at certain frequencies $\omega$, which are associated with resonances, are by definition also zero eigenstates of $J$. Usually, these are the only eigenstates of $J$ studied in mechanical systems, since the resonant modes can be directly probed.
In electric circuits, the important difference is that direct measurements via the impedance do not only involve these resonant states. From the definition $Z_{ab}=(V_a-V_b)/I$, 
the impedance measurement can be thought of as a transport problem with an arbitrarily large external driver/probe. This additional complication requires extra information from $J$, namely, the contributions from all eigenvalues of $J$, and not just the zero eigenvalue (Eq.~\ref{Z2}). 
Due to the different orders of time derivatives (powers of $\omega$) entering the equation of motion of an RLC circuit, there is no direct relation between $\omega$ and the eigenvalues of $J$. This is in contrast to mechanical systems, where the mass is local ($M$ is diagonal) and $\omega^2$ can easily be made the eigenvalue.
Still, mechanical and electrical resonances are similar in spirit despite being characterized in seemingly opposite ways. In mechanical systems, resonances are associated with minimal dissipation, where small driving perturbations can sustain large oscillations. The same is true for electrical circuits, despite having ostensibly divergent impedance: one then has large voltage oscillations corresponding to small input/output current.
Note that, barring specially constructed examples, mechanic systems generally have the restriction that the polarization directions have to be related to the relative spatial displacement of the sites; in electric circuits at laboratory scales, there is no such constraint. Our zigzag topolectrical circuits can be conveniently carried over to mechanical systems, i.e. in Ref. \onlinecite{lee2017dynamically} where its Floquet dynamics was also explored. In particular, the generalization from boundary modes to domain wall modes~\cite{Sessi1269} certainly establishes a direction worth considering in mechanical systems as well. This is less obvious for e.g. the Weyl circuit, as higher dimensional networks cannot be realized easily in a mechanical arrangement of springs.

\subsection{Microwave resonator waveguides}
The physical mechanism of microwave resonators\cite{hafezi2011robust,liang2013optical} is fundamentally different, being based on time-reversal symmetry breaking due to phases accumulated along the paths of the circuits. As such, the system is described by a unitary matrix $U=e^{iH_{\text{eff}}t}$, with the effective Hamiltonian $H_{\text{eff}}$ dependent on the phases (i.e. optical path lengths) along each edge. Formally, it maps to a Floquet Hamiltonian with quasienergies taking values on a circle. 
Such setups are also inspired by
electron dynamics through a disordered quantum Hall system.

By contrast, our RLC circuits are based purely on a direct mapping to an ordinary tight binding Hamiltonian, albeit with non-resonant modes also taking into account for handling arbitrarily large input/output currents. This is why the transfer from topolectrical circuits to mechanical systems may appear more direct than for microwave resonator arrays. This can change as one addresses Floquet phases in topolectrical circuits in the future.
\end{widetext}

\end{document}